\documentclass[aps,pra,superscriptaddress,preprint]{revtex4}
\usepackage{amsmath,epsfig,color,amssymb}
\usepackage{graphicx}

\def\be{\begin{equation}}
\def\ee{\end{equation}}
\def\bea{\begin{eqnarray}}
\def\eea{\end{eqnarray}}

\newcommand{\avg}[1]{\mbox{$\langle#1\rangle$}}

\def\omegaopt{\omega_{\footnotesize\textrm{opt}}}
\def\kappath{\kappa_{\footnotesize\textrm{th}}}
\def\bfr{{\bf r}}
\def\Uopt{U_{\footnotesize\textrm{opt}}}
\def\Umech{U_{\footnotesize\textrm{mech}}}

\begin{document}
\title{Ultrahigh-Q mechanical oscillators through optical trapping}

\date{\today}

\author{D.E. Chang}
\affiliation{Institute for Quantum Information, California Institute of Technology, Pasadena, CA 91125}
\author{K.-K. Ni}\affiliation{Norman Bridge
Laboratory of Physics 12-33, California Institute of Technology,
Pasadena, CA 91125}
\author{O. Painter}\affiliation{Thomas J. Watson, Sr., Laboratory of Applied Physics, California Institute of Technology, Pasadena, CA 91125}
\author{H.J. Kimble}\affiliation{Norman Bridge
Laboratory of Physics 12-33, California Institute of Technology,
Pasadena, CA 91125}

\begin{abstract}
Rapid advances are being made toward optically cooling a single
mode of a micro-mechanical system to its quantum ground state and
observing quantum behavior at macroscopic scales. Reaching this
regime in room-temperature environments requires a stringent
condition on the mechanical quality factor $Q_m$ and frequency
$f_m$, $Q_{m}f_{m}{\gtrsim}k_{B}T_{\footnotesize\textrm{bath}}/h$,
which so far has been marginally satisfied only in a small number
of systems. Here we propose and analyze a new class of systems
that should enable unprecedented $Q$-frequency products. The technique
is based upon using optical forces to ``trap'' and stiffen the
motion of a tethered mechanical structure, thereby freeing the
resultant mechanical frequencies and decoherence rates from
underlying material properties.

\end{abstract}
\maketitle

The coupling of a high-Q mode of a micro-mechanical oscillator to
an optical cavity has emerged as a promising route toward
observing quantum behavior at macroscopic scales~\cite{cleland09}.
This opto-mechanical interaction is being used, for example, to
optically cool a mechanical mode toward its quantum ground
state~\cite{cleland09}. Ground-state cooling requires that the
product of the quality factor $Q_m$ and frequency $f_m$ of the
mechanical mode exceed $k_{B}T_{\footnotesize\textrm{bath}}/h$,
where $h$ is Planck's constant. For a room-temperature bath, this
condition is marginally satisfied only in a small number of
current experiments~\cite{wilson09}. The ratio
$hQ_{m}f_m/k_{B}T_{\footnotesize\textrm{bath}}$ also determines
the quantum coherence time of the system relative to the
mechanical period. Significant improvements to $Q$-frequency
products are thus critical for schemes to prepare and detect
non-classical states of motion~\cite{marshall03,genes08c}.
Mechanical systems exhibiting extremely high quality factors also
offer novel opportunities for precision measurement and force
detection~\cite{geraci10,li10}.

\begin{figure}[p]
\begin{center}
\includegraphics[width=16cm]{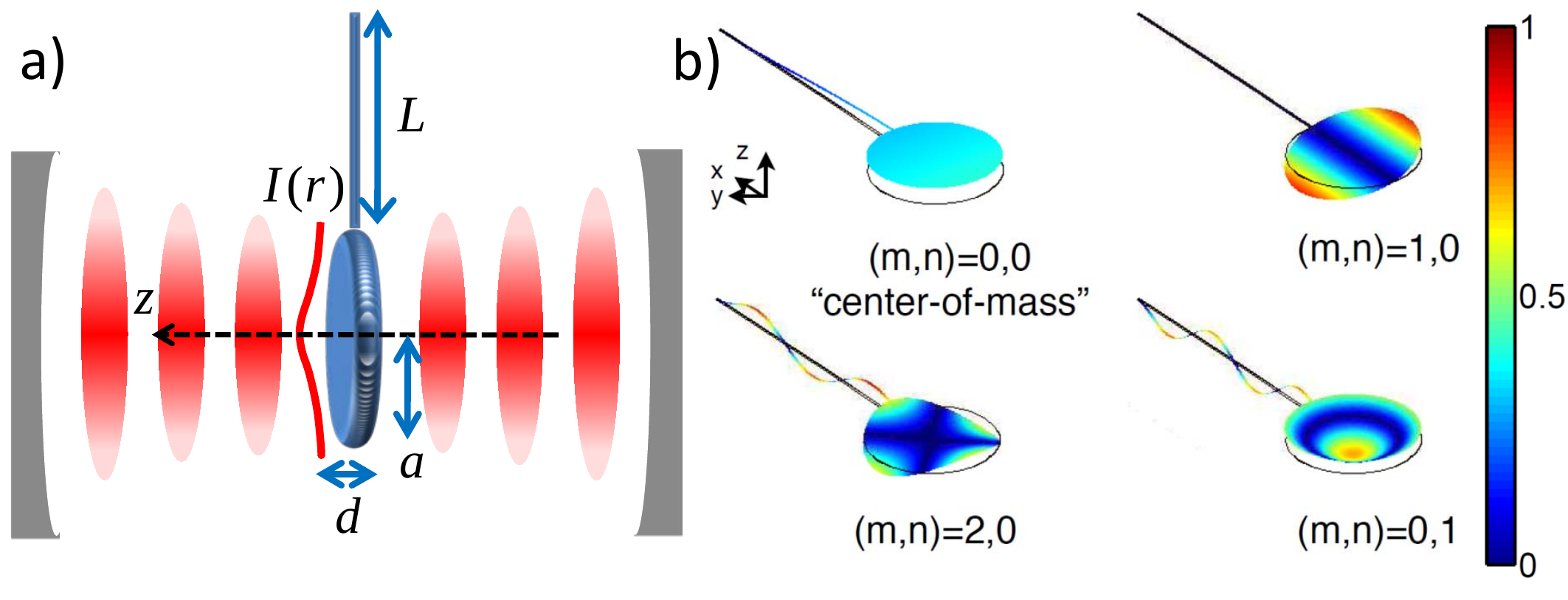}
\end{center}
\caption{a) Side view of a membrane supported by a single tether
inside a Fabry-Perot cavity. The membrane has radius $a$ and
thickness $d$, while the tether has length $L$ and a square
cross-section of width $b$. It is trapped in the anti-node of a
standing optical field with transverse intensity profile $I(r)$ at
the membrane location. b) Displacement fields of a few selected
membrane modes~(in arbitrary units), for zero trapping intensity.
The black outline indicates the equilibrium position. $(m,n)$
denote the number of nodal diameters and circles, respectively.
The system dimensions are given by $a=10\;\mu$m, $b=d=50$~nm, and
$L=50\;\mu$m.\label{fig:system}}
\end{figure}

In this Letter, we propose a class of systems that should enable
unprecedented $Q$-frequency products. The approach is based upon
optically ``trapping'' a tethered membrane with low natural
mechanical frequency in the anti-node of a strong optical standing
wave~(see Fig.~\ref{fig:system}a)~\cite{ashkin07}. While there are
many possible realizations, here we focus on a pendulum geometry,
where a relatively large disk is supported by a single thin
tether. The dielectric disk is attracted to the anti-node of the
field, leading to an optical stiffening of its flexural modes.
Under realistic conditions, the re-normalized mode frequencies can
be significantly enhanced over the values expected under material
stresses alone. Of particular interest is the
``center-of-mass''~(CM) mode, where the disk oscillates in the
optical potential with negligible flexural motion. We show that
this motion exhibits an extremely large ratio of potential energy
stored in the optical field to strain energy,
$U_{\footnotesize\textrm{opt}}/U_{\footnotesize\textrm{mech}}$.
This can yield a correspondingly large increase in the
$Q$-frequency product over a conventional mechanical system due to
the suppression of dissipation through internal friction.

Our approach to achieving long coherence times builds upon
previous proposals, which suggested that the highly isolated CM
mode of an optically levitated
nanosphere~\cite{ashkin76,libbrecht04} can enable quantum
opto-mechanics in room-temperature
environments~\cite{chang10a,romero-isart10}. Compared to the
nanosphere, our approach has two significant advantages. First,
the nanosphere scatters light omni-directionally, leading to
motional heating via photon recoil. Suppression of recoil heating
to reach the quantum regime requires spheres with sub-wavelength
volumes, $V/\lambda^3{\ll}1$~\cite{chang10a}. In contrast, the
planar membrane primarily couples the counter-propagating
components of the trapping beam, strongly reducing recoil heating
even for large systems. Second, these membranes can be fabricated
using well-established techniques that already yield excellent
mechanical and optical properties in a number of
experiments~\cite{wilson09,sankey10}. Our proposal thus shows how
the ideas of optical levitation can be brought to bear upon
``conventional'' and practically deployable mechanical systems to
yield remarkable coherence times.

We begin by considering the mechanical modes of a free, thin
circular membrane with thickness $d$ and radius $a$ in the absence
of any tethers. In equilibrium, the membrane is situated at $z=0$,
in the anti-node of an optical standing wave
$E(z)\propto\cos\,kz$~(see Fig.~\ref{fig:system}a). The optical
field polarizes the dielectric disk, yielding a gradient force
trap around $z=0$. Absent any internal forces, the optical field
traps a thin membrane~($d\ll\lambda$) with a restoring frequency
given by
$\omegaopt(r)=\left(\frac{2k^{2}I(r)(\epsilon-1)}{{\rho}c}\right)^{1/2}$,
where $r$ is the radial coordinate. Here $I(r)$ is the beam
intensity profile~(assumed to be rotationally symmetric) in the
direction transverse to $z$, $k=2\pi/\lambda$ is the optical
wavevector, $\rho$ is the mass density, and $\epsilon$ is the
dielectric constant. Now including the internal stresses, the
mechanical displacement field from equilibrium, $\zeta(x,y)$,
obeys
\be \frac{\partial^2\zeta}{{\partial}t^2}=
-\omegaopt^{2}(r)\zeta-\frac{Ed^2}{12\rho(1-\sigma^{2})}\nabla^{4}\zeta,\label{eq:waveeqn}
\ee
subject to free boundary conditions~\cite{landau86}~(also see
Appendix). $E,\sigma$ denote the Young's modulus and Poisson's
ratio, respectively. Here and in the following, $\nabla^{2}$ is
understood to be the Laplacian in the transverse plane. Due to the
rotational symmetry, we seek solutions of the form
$\zeta(x,y)=f(r)\cos\,m\theta\,e^{-i\omega_{m}t}$. The spatial
modes are indexed by the number of nodal diameters and circles,
$(m,n)$~(see Fig.~\ref{fig:system}b). For our numerical results we
take material parameters $E=270$~GPa, $\sigma=0.25$,
$\rho=2.7$~g/cm${}^3$, $\epsilon=4$ corresponding to
stoichiometric silicon nitride~\cite{wilson09}, and an operating
wavelength of $\lambda=1\;\mu$m. For a free disk without optical
forces, the fundamental $(2,0)$ flexural mode has natural
frequency
$\omega_{m,\footnotesize\textrm{nat}}^{(2,0)}/2\pi{\approx}0.25\frac{d}{a^2}\sqrt{\frac{E}{\rho(1-\sigma^2)}}$,
or
$\omega_{m,\footnotesize\textrm{nat}}^{(2,0)}/2\pi{\approx}1.3$~MHz
for a disk of dimensions $a=10\;\mu$m, $d=50$~nm. There is also a
trivial solution corresponding to the CM or $(0,0)$ mode with
constant $f(r)$ and zero frequency.

\begin{figure}[p]
\begin{center}
\includegraphics[width=17cm]{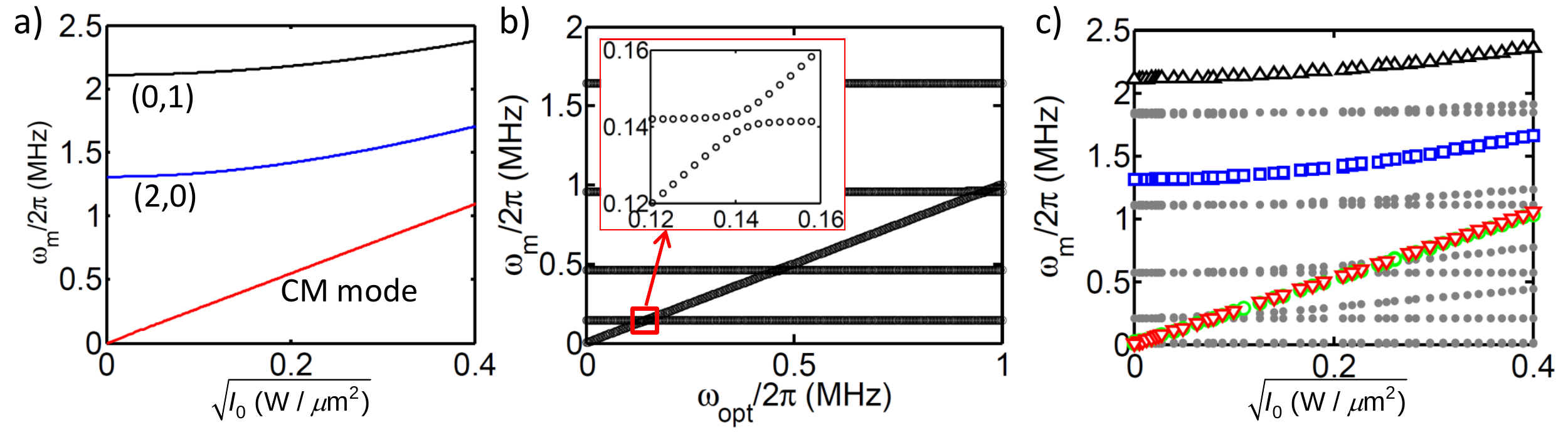}
\end{center}
\caption{a) Normal mode frequencies of a free circular disk
trapped in an optical standing plane wave, as a function of beam
intensity. The disk has a thickness and radius of $d=50$~nm and
$a=10\;\mu$m, respectively, and material properties corresponding
to stoichiometric silicon nitride. b) Normal mode frequencies of a
rigid membrane suspended by a single tether, as a function of the
optical restoring frequency $\omegaopt$ acting on the membrane.
The tether has length $L=50\;\mu$m and a square cross-section of
$b=50$~nm on each side, while the ratio of membrane to tether mass
is given by $M/m_t=125$. Away from degeneracies, the mode spectrum
consists of a CM mode with frequency ${\sim}\omegaopt$ and
discrete tether modes with frequencies $\omega_n$~($n=1,2,3,...$).
Avoided crossings occur near degeneracies
$\omegaopt{\sim}\omega_n$~(see inset). c) Normal mode frequencies
for a realistic tethered system, as a function of peak trapping
beam intensity. The disk and tether have dimensions identical to
those in a) and b), while the beam waist is $w=35\;\mu$m. The gray
points indicate tether modes. The
red~(\textcolor{red}{$\bigtriangledown$}),
green~(\textcolor{green}{$\bigcirc$}),
blue~(\textcolor{blue}{$\square$}), and black~($\bigtriangleup$)
points denote the CM, (1,0), (2,0), and (0,1) membrane modes,
respectively.\label{fig:modes}}
\end{figure}

For a uniform intensity, $I(r)=I_0$, the natural radial functions
$f^{(m,n)}_{\footnotesize\textrm{nat}}(r)$ remain eigenmodes of
Eq.~(\ref{eq:waveeqn}), but with re-normalized mechanical
frequencies given by
$\omega_{m}^{(m,n)}(I_0)=\sqrt{\omegaopt^2(I_0)+(\omega_{m,\footnotesize\textrm{nat}}^{(m,n)}){}^2}$.
Thus, the CM mode is now a non-trivial solution with frequency
$\omega_{m}^{\footnotesize\textrm{CM}}=\omegaopt$, but still
retains a uniform spatial profile. The frequencies of all the
flexural modes increase as well, with the CM mode remaining the
lowest in frequency~(see Fig.~\ref{fig:modes}a).

The absence of energy stored in internal strains for the CM motion
has the important implication of eliminating dissipation due to
internal friction. Instead, the energy is stored in an optical
potential that contributes no losses~(but can contribute a recoil
heating force, as described later). To quantify this, we consider
the influence of thermoelastic damping on our system. We focus on
thermoelastic damping because a) it can be analytically
modelled~\cite{landau86,lifshitz00}, b) it is a fundamental limit
even for ``perfect'' devices~\cite{lifshitz00}, and c) a number of
micro-mechanical systems are approaching this
limit~\cite{verbridge06,lee09}. We emphasize, however, that our
conclusions are qualitatively correct for any internal dissipative
process.

Thermoelastic damping arises because realistic materials have a
non-zero coefficient of thermal expansion. The flexural motion
creates local volume changes that then lead to temperature
gradients and heat flow. Mechanical energy must be expended to
drive this heat flow, leading to a finite $Q_m$. We make two
simplifying assumptions to the general thermoelastic
equations~\cite{landau86,lifshitz00}, which are well-justified in
our system. First, we assume that the thermoelastic coupling is
weak, so that the spatial modes determined by
Eq.~(\ref{eq:waveeqn}) are not altered to lowest order. Second,
the strains vary most rapidly along the thin direction of the
disk, and thus we ignore the relatively small transverse
temperature gradients. The temperature field is given by
$T(x,y,z,t)=T_{\footnotesize\textrm{bath}}+\Delta{T}(x,y,z,t)$,
where $\Delta{T}$ satisfies the driven heat equation
\be
\left(c_{V}\frac{\partial}{{\partial}t}-\kappath\frac{\partial^2}{{\partial}z^2}\right)\Delta{T}=\frac{E{\alpha}T_{\footnotesize\textrm{bath}}z}{3(1-2\sigma)}\frac{\partial}{\partial{t}}\nabla^{2}\zeta\label{eq:heateq}
\ee
with boundary conditions $\partial\Delta{T}/{\partial}z=0$ at
$z={\pm}d/2$. Here $c_V$ is the heat capacity per unit volume,
$\kappath$ is the thermal conductivity, and $\alpha$ is the
volumetric thermal expansion coefficient~(we take
$c_V=2$~J/cm${}^3\cdot$K, $\kappath=20$~W/m$\cdot$K,
$\alpha=4.8{\times}10^{-6}$~K${}^{-1}$ for SiN). The work done in
driving the heat flow over one cycle is
\be \Delta{W}\approx
-\frac{\kappath}{T_{\footnotesize\textrm{bath}}}\int_{0}^{2\pi/\omega_m}dt\int\,d^{3}\bfr
\Delta{T}(\bfr)(\partial^2\Delta{T}/{\partial}z^2),\label{eq:work}
\ee
and the thermoelastically limited quality factor is
$Q_{m,\footnotesize\textrm{th}}=2{\pi}(U_{\footnotesize\textrm{opt}}+U_{\footnotesize\textrm{mech}})/\Delta{W}$,
where $U_{\footnotesize\textrm{opt}}$ and
$U_{\footnotesize\textrm{mech}}$ are the energies stored in the
optical field and strains, respectively. To good approximation,
one finds that the thermoelastically limited $Q$-frequency product
is given by
$Q_{m,\footnotesize\textrm{th}}f_m=\frac{45\kappath}{{\pi}Ed^{2}T_{\footnotesize\textrm{bath}}\alpha^{2}}\frac{1-\sigma}{1+\sigma}(1+U_{\footnotesize\textrm{opt}}/U_{\footnotesize\textrm{mech}})$~(see
Appendix). Thus the storage of energy in the optical field leads
directly to an enhancement of the $Q$-frequency product. As a
useful comparison, in the absence of optical trapping, an
unstressed stoichiometric SiN film of thickness $d=50$~nm would
have a $Q$-frequency product limited to
$Q_{m,\footnotesize\textrm{th}}f_{m}{\approx}4{\times}10^{13}$~Hz
at room temperature, which only marginally exceeds the fundamental
limit
$k_{B}T_{\footnotesize\textrm{bath}}/h{\approx}6{\times}10^{12}$~Hz
needed for ground-state cooling. The $Q$-frequency product is also
proportional to the number of coherent oscillations that the
system can undergo before a single phonon is exchanged with the
thermal bath,
$N^{(\footnotesize\textrm{osc})}_{\footnotesize\textrm{th}}=Q_{m,\footnotesize\textrm{th}}f_{m}h/(2{\pi}k_{B}T_{\footnotesize\textrm{bath}})$~($N^{(\footnotesize\textrm{osc})}_{\footnotesize\textrm{th}}{\sim}1$
for the conventional membrane described above). A large value is
critical to preparing and observing quantum superposition or
entangled states~\cite{marshall03,genes08c}.

We now examine thermoelastic damping of our free disk. Clearly, if
$\omegaopt(r)$ is spatially uniform, the CM mode has no internal
strains and experiences zero thermoelastic dissipation. A finite
beam waist creates inhomogeneous optical forces that internal
stresses must compensate for, which mixes CM and internal motion
together~(we still refer to this mixed mode as the ``CM''). The
mixing becomes significant when the variation in $\omegaopt(r)$
across the disk overtakes the natural rigidity of the system~(as
characterized by the natural fundamental frequency), and can be
avoided by using sufficiently large beam waists or low
intensities.  In this regime, significant enhancements to the
$Q$-frequency product should result.

\begin{figure}[p]
\begin{center}
\includegraphics[width=17cm]{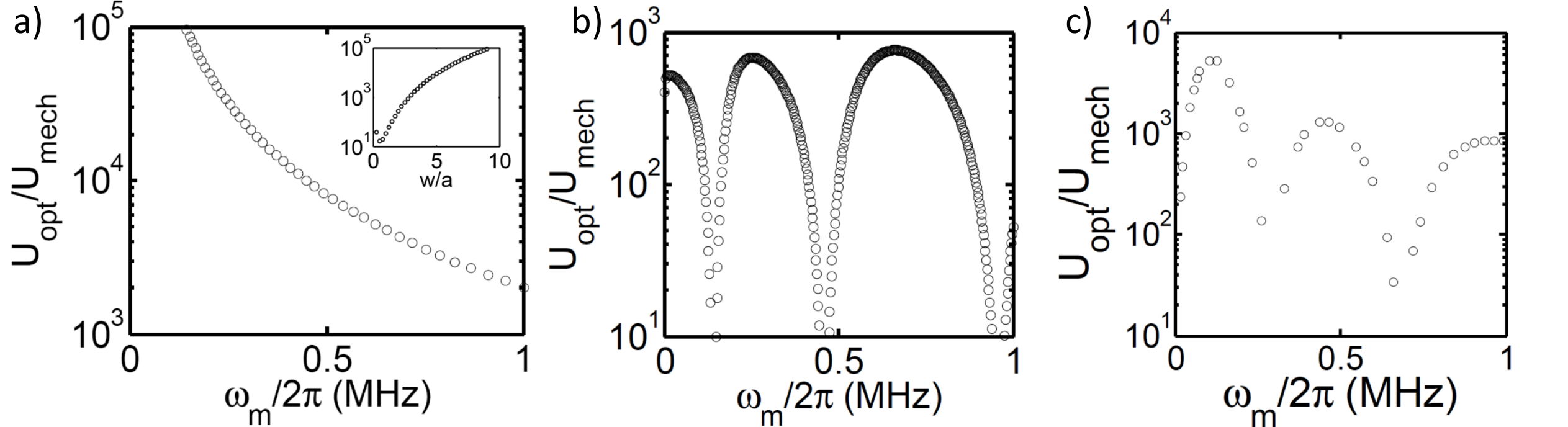}
\end{center}
\caption{Ratio
$U_{\footnotesize\textrm{opt}}/U_{\footnotesize\textrm{mech}}$ of
optical to strain energy for the CM mode of different systems: a)
an optically trapped free disk as a function of its frequency
$\omega_m/2\pi$. Here the beam waist is fixed at $w=35\;\mu$m and
the trap intensity is varied to yield the corresponding
$\omega_m$. Inset: the trap intensity is fixed such that
$\omega_{m}/2\pi=1$~MHz in the plane wave limit while $w/a$ is
varied. b) a rigid membrane suspended by a single tether, as a
function of the CM frequency. c) a realistic tethered structure,
obtained by finite-element simulations. The system dimensions for
these plots are identical to those in Fig.~\ref{fig:modes},
namely, $a=10\;\mu$m, $L=50\;\mu$m,
$d=b=50$~nm.\label{fig:thermoelastic}}
\end{figure}

These effects are illustrated in Fig.~\ref{fig:thermoelastic}a.
For concreteness, we assume that the trapping beam has a Gaussian
profile with waist $w$, $I(r)=I_{0}e^{-2r^2/w^2}$~(for now, we
ignore possible corrections due to distortion as the beam
diffracts around the disk). In Fig.~\ref{fig:thermoelastic}a, we
plot the ratio
$U_{\footnotesize\textrm{opt}}/U_{\footnotesize\textrm{mech}}$ for
the CM mode as a function of its frequency, which is varied
through the peak intensity $I_0$. We use the same disk dimensions
as before and a waist of $w=35\;\mu$m. The energy ratio
monotonically decreases, reflecting the increased inhomogeneity in
the optical potential. In the inset, a complementary process is
illustrated, where the beam waist $w$ is varied, while the peak
intensity $I_0$ is fixed such that $\omegaopt(I_0)/2\pi=1$~MHz.
The energy ratio increases indefinitely with $w/a$ and approaches
infinity in the plane-wave limit.

We now consider the realistic pendulum geometry shown in
Fig.~\ref{fig:system}a, where the tether provides an extremely
weak restoring force for the ``CM'' motion of the disk. We first
present a simplified analysis that isolates the role of the tether
on the mode spectrum and $Q$-frequency product. Specifically, we
treat the membrane as a perfectly rigid point particle of mass
$M$, which experiences an optical restoring force with frequency
$\omega_{\footnotesize\textrm{opt}}$, while internal stresses
alone act on the tether. Then, for a tether of length $L$ whose
long axis is situated along $x$, the displacement field
$\phi(x,t)$~(where $0 \leq x \leq L$) satisfies the beam
equation~\cite{landau86},
\be \frac{\partial^{2}\phi}{\partial
t^2}=-\frac{Eb^2}{12\rho}\frac{\partial^{4}\phi}{\partial x^4}.
\ee
Here $b$ denotes the width of the tether~(assumed to be square in
cross-section). The beam is clamped at $x=0$,
$\phi(0,t)=\partial_{x}\phi(0,t)=0$, while at $x=L$ the boundary
conditions are given by $\partial_{x}^{2}\phi(L,t)=0$ and
$M\partial_{t}^2\phi(L,t)=-M\omegaopt^2\phi(L,t)+Eb^4\partial_{x}^3\phi(L,t)/12$.
The last equation describes the acceleration of the membrane due
to optical restoring forces and the shear force imparted by the
tether.

It is straightforward to solve for the system eigenmodes~(see
Appendix) and the results are summarized here. For large mass
ratios between the membrane and tether,
$M/m_t{\rightarrow}\infty$, the modes usually consist of a CM mode
for the membrane with frequency
$\omega_{m}^{\footnotesize\textrm{CM}}\approx\sqrt{\omega_p^2+\omegaopt^2}$
and a set of discrete tether modes with frequencies
$\omega_n\approx\left(\frac{(n+1/4)\pi}{{\beta}L}\right)^2$, where
$\beta=(12\rho/Eb^2)^{1/4}$. The CM mode spectrum is understood as
a low-frequency ``pendulum'' mode~(with natural frequency
$\omega_p\approx\sqrt{Eb^{4}/4ML^3}$, where
$\omega_p{\ll}\omegaopt,\omega_n$ for our systems of interest)
whose frequency can be strongly re-normalized by the optical
force, while the tether mode spectrum results from the heavy
membrane essentially acting as a second clamp. This description
holds except near degeneracies $\omegaopt{\approx}\omega_n$, where
coupling between the tether and membrane motions yields an avoided
crossing whose width decreases with increasing mass ratio $M/m_t$.
This result is illustrated in Fig.~\ref{fig:modes}b, for a mass
ratio of $M/m_t=125$~(corresponding to the disk size considered
earlier attached to a tether of length $L=50\;\mu$m and width
$b=50$~nm). In Fig.~\ref{fig:thermoelastic}b, we plot the energy
ratios
$U_{\footnotesize\textrm{opt}}/U_{\footnotesize\textrm{mech}}$ for
the CM motion as a function of $\omegaopt$. Here, the strain
energy is completely attributable to the tether, as we take the
membrane to be a rigid object. The energy ratio is dramatically
reduced near the avoided crossings, while away from these
crossings, the energy ratio plateaus to a value near
$U_{\footnotesize\textrm{opt}}/U_{\footnotesize\textrm{mech}}{\sim}8M/m_t$~(see
Appendix).

For a realistic tethered system~(as in Fig.~\ref{fig:system}b)
where the membrane is not perfectly rigid, mode mixing between the
tether and membrane and mixing between the CM and internal
membrane motion will occur simultaneously. We have numerically
solved the full stress-strain equations for such a system using
COMSOL, a commercial finite-element simulation package. A
characteristic mode spectrum is plotted in Fig.~\ref{fig:modes}c
as a function of the peak trapping intensity $I_0$, for parameters
$a=10\;\mu$m, $L=50\;\mu$m, $d=b=50$~nm, and $w=35\;\mu$m. Away
from avoided crossings, the modes can clearly be characterized as
tether modes~(gray points) or membrane modes~(color). For our
particular choice of beam waist size, the tethers themselves
experience optical restoring forces, leading to a slight optical
stiffening of tether modes that have displacements along the
optical propagation axis. Comparing the membrane modes, the CM
mode lies lower in frequency than the flexural modes, as in the
case of a free disk. A nearly degenerate torsional $(1,0)$ mode
also exists, which in principle should have no opto-mechanical
coupling to the cavity field and can be ignored~(see Appendix).

In Fig.~\ref{fig:thermoelastic}c, we plot the energy ratio
$U_{\footnotesize\textrm{opt}}/U_{\footnotesize\textrm{mech}}$ for
the CM mode of our tethered structure as a function of its
frequency. The features displayed here are clearly a combination
of those appearing in the limiting cases of a free disk and a
rigid disk attached to a tether. In particular, large plateaus in
$U_{\footnotesize\textrm{opt}}/U_{\footnotesize\textrm{mech}}$
appear away from avoided crossings with tether modes. The plateau
heights are higher than the limit
$U_{\footnotesize\textrm{opt}}/U_{\footnotesize\textrm{mech}}{\sim}8M/m_t$
predicted by the simple model, since the tether experiences an
optical spring force as well. The decrease in the plateau heights
with increasing CM frequency is associated with increased mixing
between pure CM and internal membrane motion. For the realistic
geometry considered here, an enhancement in the $Q$-frequency
product on the order of ${\sim}10^3$ compared to a conventional
system can be realized at a frequency of
$\omega_m/2\pi{\sim}1$~MHz. For a thermoelastically limited
system, this corresponds to a coherence time of
$N^{(\footnotesize\textrm{osc})}_{\footnotesize\textrm{th}}{\sim}10^3$.

Thus far, we have assumed that the trapping beam has a Gaussian
profile. For cavity opto-mechanics~\cite{cleland09}, it will be
necessary to trap the membrane within a Fabry-Perot cavity, as
illustrated in Fig.~\ref{fig:system}a. For example, here, a
relatively strong beam could be used for trapping, while a second,
weaker beam with a non-zero intensity gradient at the trap
position would facilitate cooling of the CM motion or quantum
state transfer processes~\cite{chang10a,romero-isart10}. The
membrane scatters and diffracts the cavity light, which introduces
two important effects. First, the mode will no longer be Gaussian,
and the new optical mode accommodated by the cavity mirrors and
the corresponding optical forces must be determined. Second,
photon scattering out of the cavity reduces cavity finesse, and
the associated random momentum kicks~(``photon recoil'') imparted
on the membrane lead to additional decoherence. We have performed
detailed simulations of the modified cavity fields and coupled
them to the equation of motion~(\ref{eq:waveeqn}) for the
membrane~(see Appendix). We find that for realistic cavity
geometries, a value of
$N^{(\footnotesize\textrm{osc})}_{\footnotesize\textrm{tot}}{\sim}10^3$
in a room-temperature environment can still be obtained, taking
into account both thermoelastic processes and recoil heating.
Remarkably, the coherence time of this system is comparable to
that of a much smaller levitated nanosphere of radius
$r{\sim}25$~nm~\cite{chang10a}, or a conventional $1$~MHz
oscillator with $Q_m$ exceeding $10^8$ at a bath temperature of
$1$~K.

We emphasize that our approach to reaching the quantum regime is
fundamentally different than other ``optical spring'' proposals
based upon optical backaction forces~\cite{braginsky02,corbitt07}.
In the latter case, the linear coupling of the mechanical
displacement to the intra-cavity intensity can yield a dynamic
optical spring effect. This effect, however, is accompanied by
significant Raman scattering of the optical pump field, which
causes phonons to be rapidly removed and added to the system.
While this does not preclude ground state
cooling~\cite{corbitt07}, a more detailed analysis~(see Appendix)
shows that it imposes severe limitations on the quantum coherence
time and makes it difficult to prepare, \textit{e.g.}, quantum
superposition states. In contrast, in our scheme, the phonons are
truly long-lived excitations. Furthermore, regardless of the
trapping scheme used, our analysis properly captures the role that
strong, spatially non-uniform optical forces have in mixing
internal motion, which is neglected in lowest-order
opto-mechanical models but relevant to most flexural systems~(see
Appendix).

Although we have focused on thermoelastic losses in the above
calculations, we expect similar improvements for any other
internal damping mechanism. The key idea is that it is possible to
circumvent natural material limits of damping by storing energy in
a lossless optical field rather than the internal strain. By
making the ratio of these energies large,
$U_{\footnotesize\textrm{opt}}/U_{\footnotesize\textrm{mech}}{\gg}1$,
any internal losses can be suppressed by a corresponding degree.
This fundamental observation allows one to design a novel class of
mechanical systems that can be fabricated and deployed using
conventional techniques, yet yield $Q$-frequency products that are
several orders of magnitude higher than previous systems. We
believe that this work will stimulate further investigation into
the relationship between optical forces and material dissipation
in a number of systems where the mechanical motion can be strongly
renormalized by light~\cite{corbitt07,rosenberg09}. Furthermore,
we anticipate that such studies will open up interesting
possibilities for quantum manipulation of mechanical systems in
room-temperature environments.

The authors thank Dal Wilson and Richard Norte for many helpful
discussions. DEC acknowledges support from the NSF~(Grant No.
PHY-0803371) and the Gordon and Betty Moore Foundation through
Caltech's Center for the Physics of Information~(CPI). KN
acknowledges support from the CPI. HJK and OJP acknowledge support
from the DARPA ORCHID program. HJK also acknowledges support from
the NSF and DoD NSSEFF.

\appendix

\section{Equation of motion for a free disk}

Here we derive the equation of motion for a thin non-uniform disk
of thickness $d(r)$, which has a reflection symmetry around
$z=0$~(such that the surface of the plate is located at
$z={\pm}d(r)/2$). We are interested in the situation where the
thickness is much less than the characteristic transverse
size~(\textit{e.g.}, the radius $a$ of a circular disk), such that
its degree of freedom along the thin direction can be effectively
eliminated and the flexural motion can be described by a
two-dimensional displacement field $\zeta(x,y)$. The equation of
motion for $\zeta(x,y)$ can be obtained by a generalization of the
derivation for a uniform disk given in Ref.~\cite{landau86}.
Specifically, the energy associated with the displacement field
$\zeta(x,y)$ is given by
\be
\Umech=\frac{E}{24(1-\sigma^2)}\int\,dx\,dy\,d(x,y)^3\left[\left(\zeta_{xx}+\zeta_{yy}\right)^2+2(1-\sigma)\left((\zeta_{xy})^2-\zeta_{xx}\zeta_{yy}\right)\right].\label{eq:F}
\ee
Here $E,\sigma$ are the Young's modulus and Poisson's ratio,
respectively, while
$\zeta_{xx}=\frac{\partial^{2}\zeta}{{\partial}x^2}$, etc. To
derive the equilibrium field $\zeta(x,y)$ under some external
normal pressure $P(x,y)$, we employ the variational principle to
minimize the system energy. Under small variations $\delta\zeta$,
and following some algebra, the variation in $\Umech$ can be
written as the sum of an integral over the transverse area of the
disk and two integrals over the circumference or edge of the disk,
\be
{\delta}\Umech=\int\,dx\,dy\,Z_{1}\delta\zeta+\oint\,dl\,Z_{2}\frac{\partial\delta\zeta}{{\partial}n}+\oint\,dl\,\,Z_{3}\delta\zeta.
\ee
Here $Z_i$ are complicated expressions involving $\zeta$ and
$d(r)$ whose forms are given below, while $n$ denotes the normal
to the edge of the disk. The integral over the disk area yields
the equilibrium equation of the disk, $Z_1=P(x,y)$, or the
dynamical equation can be obtained by replacing
$P(x,y){\rightarrow}-{\rho}d(r)\frac{\partial^2\zeta}{{\partial}t^2}$.
Doing so, and including the effect of external optical forces, one
finds
\be \frac{\partial^2\zeta}{{\partial}t^2} =
-\omegaopt^{2}(r)\zeta-\frac{E}{12\rho(1-\sigma^{2})d(r)}\left(\nabla^{2}\left(g(r)\nabla^{2}\zeta\right)
-(1-\sigma)\left(\zeta_{yy}g_{xx}+\zeta_{xx}g_{yy}-2\zeta_{xy}g_{xy}\right)\right).\label{eq:waveeqn}
\ee
Here we have defined $g(r)=d(r)^3$, and $\nabla^{2}$ is understood
to be the Laplacian in the transverse plane. As described in the
main text,
$\omegaopt(r)=\left(\frac{2k^{2}I(r)(\epsilon-1)}{{\rho}c}\right)^{1/2}$.
For a disk with free boundary conditions at the edge, the
quantities $\delta\zeta$ and $\partial\delta\zeta/{\partial}n$ are
arbitrary on the boundary, so the coefficients $Z_{2,3}$ should
vanish, yielding the two boundary conditions. Defining $n$ and $l$
to be the normal and tangential directions to the edge of the
disk, and $\theta$ as the local angle between $x$ and $n$, these
boundary conditions become
\bea 0 & = &
\frac{\partial}{{\partial}n}\left(g\nabla^{2}\zeta\right)+(1-\sigma)\left[\frac{\partial}{{\partial}l}\left(-g\zeta_{xy}\cos\,2\theta+(g/2)\sin\,2\theta(\zeta_{xx}-\zeta_{yy})\right)\right.
\nonumber \\ & &
+\left.\cos\theta(\zeta_{yy}g_{x}-\zeta_{xy}g_{y})+\sin\theta(\zeta_{xx}g_{y}-\zeta_{xy}g_{x})\right],
\nonumber \\ 0 & = & \nabla^{2}\zeta +(1-\sigma)
\left(2\zeta_{xy}\sin\theta\cos\theta-\zeta_{yy}\cos^{2}\theta-\zeta_{xx}\sin^{2}\theta\right).\label{eq:bc}
\eea
%

\section{Thermoelastic damping}

We begin with Eqs.~(2) and~(3) in the main text that describe
thermoelastic damping,
\be
\left(c_{V}\frac{\partial}{{\partial}t}-\kappath\frac{\partial^2}{{\partial}z^2}\right)\Delta{T}=\frac{E{\alpha}T_{\footnotesize\textrm{bath}}z}{3(1-2\sigma)}\frac{\partial}{\partial{t}}\nabla^{2}\zeta\label{eq:heateq}
\ee
and
\be \Delta{W}\approx
-\frac{\kappath}{T_{\footnotesize\textrm{bath}}}\int_{0}^{2\pi/\omega_m}dt\int\,d^{3}\bfr
\Delta{T}(\bfr)(\partial^2\Delta{T}/{\partial}z^2).\label{eq:work}
\ee
The first equation describes the driven heat equation along the
thin direction of the membrane, with boundary conditions
$\partial\Delta{T}/{\partial}z=0$ at $z={\pm}d(r)/2$. Here $c_V$
is the heat capacity per unit volume, $\kappath$ is the thermal
conductivity, and $\alpha$ is the volumetric thermal expansion
coefficient~(we take $c_V=2$~J/cm${}^3\cdot$K,
$\kappath=20$~W/m$\cdot$K, $\alpha=4.8{\times}10^{-6}$~K${}^{-1}$
for SiN). The second equation describes the amount of work done in
driving the heat flow over one cycle for a particular mechanical
eigenmode. Solving Eq.~(\ref{eq:heateq}) and substituting into
Eq.~(\ref{eq:work}), one finds
\be \Delta{W}\approx\frac{\pi\omega_m\alpha^2 E^2 d^5
T_{\footnotesize\textrm{bath}}}{1080\kappath(1-\sigma)^2}\int\,dx\,dy\,(\nabla^{2}\zeta)^2.
\ee

Let us now compare this quantity with the total strain energy
$\Umech$ given in Eq.~(\ref{eq:F}). For simplicity, here we
specialize to the case where the disk has a uniform thickness $d$,
such that
\be
\Umech=\frac{Ed^3}{24(1-\sigma^2)}\int\,dx\,dy\,\left[(\nabla^{2}\zeta)^2+2(1-\sigma)\left((\zeta_{xy})^2-\zeta_{xx}\zeta_{yy}\right)\right].
\ee
Note that $\Delta{W}$ and $\Umech$ have similar forms, as both
involve an integral over the membrane area of the quantity
$(\nabla^2\zeta)^2$. The strain energy contains a second term,
however, whose relative importance we characterize now. The second
term can in fact be re-written as a line integral around the
circumference of the membrane,
\be
\int\,dx\,dy\,(\zeta_{xy})^2-\zeta_{xx}\zeta_{yy}=\oint\,dx\,\zeta_{xy}\zeta_{x}-\oint\,dy\,\zeta_{yy}\zeta_{x}.
\ee
This boundary integral vanishes identically for certain types of
shapes or boundary conditions, such as a clamped membrane. For our
free disk, this boundary term does not identically vanish, but
numerically we can confirm that the boundary contribution is small
relative to the total strain energy. To good approximation then,
we can write
$\Umech\approx\frac{Ed^3}{24(1-\sigma^2)}\int\,dx\,dy\,(\nabla^{2}\zeta)^2$.
This leads to the expression for the thermoelastically limited
$Q$-frequency product given in the main text,
\be
Q_{m,\footnotesize\textrm{th}}f_m=\frac{45\kappath}{{\pi}Ed^{2}T_{\footnotesize\textrm{bath}}\alpha^{2}}\frac{1-\sigma}{1+\sigma}\left(1+\frac{\Uopt}{\Umech}\right).
\ee
%

\section{Rigid membrane attached to tether}

Here, we derive in detail the properties of a rigid membrane of
mass $M$ attached to a single tether. As described in the main
text, the tether has a length $L$~(oriented along the $x$-axis)
and a square cross-section of width $b$. The displacement field
$\phi(x,t)$~(where $0 \leq x \leq L$) satisfies the beam
equation~\cite{landau86},
\be \frac{\partial^{2}\phi}{\partial
t^2}=-\frac{Eb^2}{12\rho}\frac{\partial^{4}\phi}{\partial
x^4}.\label{eq:beameqn} \ee
The beam is clamped at $x=0$, $\phi(0,t)=\partial_{x}\phi(0,t)=0$,
while at $x=L$ the boundary conditions are given by
$\partial_{xx}\phi(L,t)=0$ and the force equation of the membrane
attached there,
$M\partial_{t}^2\phi(L,t)=-M\omegaopt^2\phi(L,t)+Eb^4\partial_{xxx}\phi(L,t)/12$.
The last equation describes the acceleration of the membrane due
to optical restoring forces and the shear force imparted by the
tether.

The boundary conditions at $x=0$ require that the general
solutions of Eq.~(\ref{eq:beameqn}) take the form
\be \phi(x)=c_{1}(\sin\,kz-\sinh\,kz)+c_{2}(\cos\,kz-\cosh\,kz),
\ee
where the dispersion relation is given by $\omega=(k/\beta)^2$ and
$\beta=(12\rho/Eb^2)^{1/4}$. The two boundary conditions at $x=L$
can be written in matrix form as $Q(c_1\;c_2)^{T}=0$, and the
corresponding equation for the eigenfrequencies,
$\textrm{det}\;Q=0$, reads
\be
M(\omega^2-\omegaopt^2)(\cos\gamma\sinh\gamma-\sin\gamma\cosh\gamma)+\frac{Eb^4\beta^{3}\omega^{3/2}}{12}(1+\cos\gamma\cosh\gamma)=0,\label{eq:beameigs}
\ee
where $\gamma=\beta L\sqrt{\omega}$. With no optical
forces~($\omegaopt=0$) and large membrane to tether mass ratio
$M/m_t{\gg}1$, the solutions to the above mode equation consist of
a low-frequency ``pendulum mode'' with frequency
$\omega_p\approx\sqrt{Eb^{4}/4ML^3}$ and a set of discrete tether
oscillation modes with frequencies $\omega_n$, which approximately
satisfy the relation
$\cos\gamma\sinh\gamma-\sin\gamma\cosh\gamma=0$. For sufficiently
large $n$~(\textit{i.e.}, large enough $\gamma$), the natural
tether frequencies asymptotically approach
$\omega_{n}=({\pi}n+\pi/4)^2/(\beta L)^2$. With optical forces, a
power series expansion of Eq.~(\ref{eq:beameigs}) reveals that the
pendulum mode frequency becomes re-normalized to the value
$\omega_{m}^{\footnotesize\textrm{CM}}=\sqrt{\omega_p^2+\omegaopt^2}$,
which can be associated with the CM mode.

Generically, for sufficiently large optical forces
$\omegaopt\gg\omega_p$, a large membrane mass $M$ means that the
eigenvalue condition of Eq.~(\ref{eq:beameigs}) is approximately
satisfied when
$M(\omega^2-\omegaopt^2)(\cos\gamma\sinh\gamma-\sin\gamma\cosh\gamma)\approx
0$, which means that the solutions typically consist of a single
CM mode with frequency $\omega_m{\approx}\omegaopt$ or tether
modes satisfying
$\cos\gamma\sinh\gamma-\sin\gamma\cosh\gamma\approx 0$. When both
conditions are satisfied simultaneously~(\textit{e.g.}, near a
degeneracy point), the second term of Eq.~(\ref{eq:beameigs}) must
be taken into account, which yields avoided crossings that mix the
tether and CM modes together.

We now consider the ratio of energy stored in the optical field to
the strain energy,
$U_{\footnotesize\textrm{opt}}/U_{\footnotesize\textrm{mech}}$,
for the CM motion away from an avoided crossing. The energy stored
in the optical field for this system is simply given by
$U_{\footnotesize\textrm{opt}}=(1/2)M\omegaopt^2\phi(L)^2$, while
the strain energy in the beam is given by~\cite{landau86}
\be
U_{\footnotesize\textrm{mech}}=\frac{Eb^4}{24}\int_{0}^{L}dx\,(\phi_{xx})^2.
\ee
It can readily be shown that for frequencies where $\gamma{\gg}1$,
the spatial modes are given to good approximation by
\be \phi(x){\approx}c_{1}\left(\sin\,kx-\cos\,kx+e^{-kx}\right).
\ee
Then, when the CM mode is positioned halfway in between two tether
modes~(say at a frequency
$\omega_m^{\footnotesize\textrm{CM}}{\approx}({\pi}n+3\pi/4)^2/(\beta
L)^2$), evaluation of the stored energies yields
$U_{\footnotesize\textrm{opt}}/U_{\footnotesize\textrm{mech}}{\sim}8M/m_t$.

\section{Dynamic optical spring}

An increase in the frequency of a mechanical mode can also be
achieved through a dynamic backaction effect in an opto-mechanical
system~\cite{braginsky02,corbitt07}, as opposed to the ``static''
optical potential considered in our work. We briefly describe the
dynamic effect and compare the two mechanisms here. Specifically,
we consider a mechanical degree of freedom whose displacement is
linearly coupled to the optical cavity frequency. The
corresponding Hamiltonian for such a system in a rotating frame
is~\cite{marquardt07,wilson-rae07}
\be
H_{int}=\frac{\hat{p}^2}{2m}+\frac{1}{2}m\omega_m^{2}\hat{z}^2-\hbar\delta\hat{a}^{\dagger}\hat{a}-{\hbar}\omega'\hat{z}\hat{a}^{\dagger}\hat{a}-\hbar\Omega_L(\hat{a}+\hat{a}^{\dagger}).\label{eq:Hom}
\ee
Here $\hat{a}$ is the annihilation operator for the optical mode,
$\hat{z},\hat{p}$ are the position and momentum operators
corresponding to the mechanical resonator, $\omega_m$ is the
natural mechanical frequency, $\delta=\omega_{L}-\omega_0$ is the
frequency detuning between an external pump field driving the
optical cavity and the optical resonance frequency
$\omega_{0}$~(when the mechanical resonator is in equilibrium),
$\Omega_L$ is the driving amplitude, and $\omega'$ is the optical
cavity frequency shift per unit mechanical displacement. In
addition to the Hamiltonian terms, the optical cavity is assumed
to have losses characterized by a linewidth $\kappa$.

For weak opto-mechanical coupling, it is customary to linearize
the optical cavity dynamics around the classical steady-state
value $\avg{\hat{a}}=\alpha=i\Omega_L/(\kappa/2-i\delta)$~(here we
have incorporated a steady-state shift of the optical resonance
frequency into our definition of the detuning $\delta$), and
eliminate the cavity to yield an effective susceptibility
$\chi(\omega)$ of the mechanical displacement in response to an
external force $f(\omega)$~\cite{marquardt07}. Specifically, one
finds
\be
\chi(\omega)^{-1}=\omega_m^2-\omega^2+\frac{16\omega_{m}\delta\Omega_m^2}{4\delta^2+(\kappa-2i\omega)^2}.
\ee
Here we have defined an effective opto-mechanical driving
amplitude $\Omega_m=g\alpha$, and
$g=\omega'z_{zp}=\omega'\sqrt{\frac{\hbar}{2m\omega_m}}$ is the
optical cavity frequency shift per unit mechanical zero-point
uncertainty. In the perturbative limit, this expression can be
written in terms of the susceptibility of a simple oscillator with
an effective linewidth and frequency that is modified due to
opto-mechanical interactions,
$\chi(\omega)^{-1}\approx\omega_{m,\footnotesize\textrm{eff}}^2-\omega^2-i\omega\Gamma_{\footnotesize\textrm{eff}}$.
The effective linewidth and mechanical frequency shift can be
interpreted as resulting from optically-induced cooling~(or
heating) and a dynamic optical spring constant, respectively. In
the relevant regime of large detuning $\delta{\gg}\omega,\kappa$,
and when the optical spring is dominant compared to the natural
mechanical frequency, the effective mechanical frequency is given
by
\be
\omega_{m,\footnotesize\textrm{eff}}{\approx}2\Omega_m\sqrt{\frac{\omega_m}{\delta}}.
\ee
The damping rate is given by
\be
\Gamma_{\footnotesize\textrm{eff}}\approx\Omega_m^2\kappa\frac{\omega_m}{\omega_{m,\footnotesize\textrm{eff}}}\left[\frac{1}{(\kappa/2)^2+(\delta+\omega_{m,\footnotesize\textrm{eff}})^2}-\frac{1}{(\kappa/2)^2+(\delta-\omega_{m,\footnotesize\textrm{eff}})^2}\right],
\ee
which is interpreted as the difference between anti-Stokes and
Stokes scattering rates. Note that for positive detuning
$\delta>0$, the opto-mechanical interaction yields an increase in
the mechanical frequency but an anti-damping force
($\Gamma_{\footnotesize\textrm{eff}}<0$). One can achieve
simultaneous stiffening and cooling by employing multiple beams
with different amplitudes and detunings~\cite{corbitt07}, but for
our purposes it is sufficient to consider only the beam that leads
to stiffening.

We wish to consider the ratio of the effective mechanical
frequency to the rate of decoherence $\Gamma_d$ induced by optical
Raman scattering, which is given by the sum of the anti-Stokes and
Stokes scattering rates. In the relevant limit of large detuning
and dominant optical spring effect, one finds
\be
\frac{\omega_{m,\footnotesize\textrm{eff}}}{\Gamma_d}{\approx}\frac{2\delta}{\kappa}.
\ee
This result states that the cavity must be driven very far off
resonance in order to yield a frequency shift that is much larger
than the decoherence rate. Operating at large detuning in turn
requires extremely large cavity input powers to get an appreciable
optical spring effect. As an example, we consider the dynamic
spring constant for a realistic geometry, such as a Fabry-Perot
cavity of length $L=1$~cm and cavity finesse
$\mathcal{F}=10^5$~(with the cavity linewidth given by
$\kappa={\pi}c/(\mathcal{F}L)=2\pi{\times}150$~kHz for our
specific parameters). The optical driving amplitude is related to
the input power $P_i$ through
$\Omega_L=\sqrt{{\kappa}P_i/(2\hbar\omega_L)}$ for perfect
in-coupling efficiency, while the opto-mechanical interaction
strength is of order $\omega'{\sim}\omega_{0}/L$. The operating
wavelength is taken to be $\lambda=1\;\mu$m. We also assume that
the SiN membrane has a radius of radius $a=10\;\mu$m and that it
undergoes pure CM motion~(such that its effective motional mass is
the same as the physical mass). Then, an input power of
$P_i{\sim}2$~kW is required if one wants to achieve a number of
coherent oscillations
$N^{(\footnotesize\textrm{osc})}=\frac{\omega_{m,\footnotesize\textrm{eff}}}{2\pi\Gamma_d}{\sim}10^3$
and an effective mechanical frequency of
$\omega_{m,\footnotesize\textrm{eff}}{\sim}2\pi\times$1~MHz. This
corresponds to an input intensity of ${\sim}10$~W/$\mu$m${}^2$ for
a beam focused to a size comparable to the membrane radius.

In contrast, in our static trapping scheme, a comparable
mechanical frequency and coherence time can be achieved for an
\textit{intra-cavity} intensity of ${\sim}0.1$~W/$\mu$m${}^2$, and
the cavity can be driven resonantly to facilitate the intra-cavity
field buildup. Because the static trap results in the membrane
being trapped at an anti-node, there is no linear opto-mechanical
coupling for the trapping field and the lowest-order
opto-mechanical coupling is quadratic in nature. The decoherence
rates caused by anti-Stokes and Stokes scattering in this case~(at
frequencies $\omega_{L}{\pm}2\omega_m$) have been calculated in
Ref.~\cite{nunnenkamp10} and are extremely rare for our realistic
systems~(occurring at a sub-Hz level). In fact, the dominant
decoherence mechanism arising from the trapping beam is due to
scattering out of the cavity and the photon recoil heating
imparted on the membrane, as described later.

\section{Modification of opto-mechanical coupling strengths}

Our theory of optical trapping of membranes predicts dramatic
corrections to the simple model of opto-mechanical interactions
given by Eq.~(\ref{eq:Hom}), when the optical restoring
forces~(either static or dynamic) become large compared to the
natural ridigity of the membrane. In this scenario of strong
optical forces, the mechanical mode shape and thus the
opto-mechanical coupling strength $g$ become functions of
intensity as well, with $g$ generally decreasing with larger
intensity.  The origin of this effect is intuitively seen by
considering a membrane that interacts with a Gaussian cavity mode
whose beam waist $w$ is smaller than the membrane radius $a$.
Then, if the optical restoring forces are large compared to the
membrane stiffness, the optical beam in fact resembles a new
boundary condition that ``pins'' the region $r\lesssim w$ of the
membrane into place. This reduces the overlap between the
mechanical displacement field and the optical beam, and thus $g$.

\begin{figure}[p]
\begin{center}
\includegraphics[width=12cm]{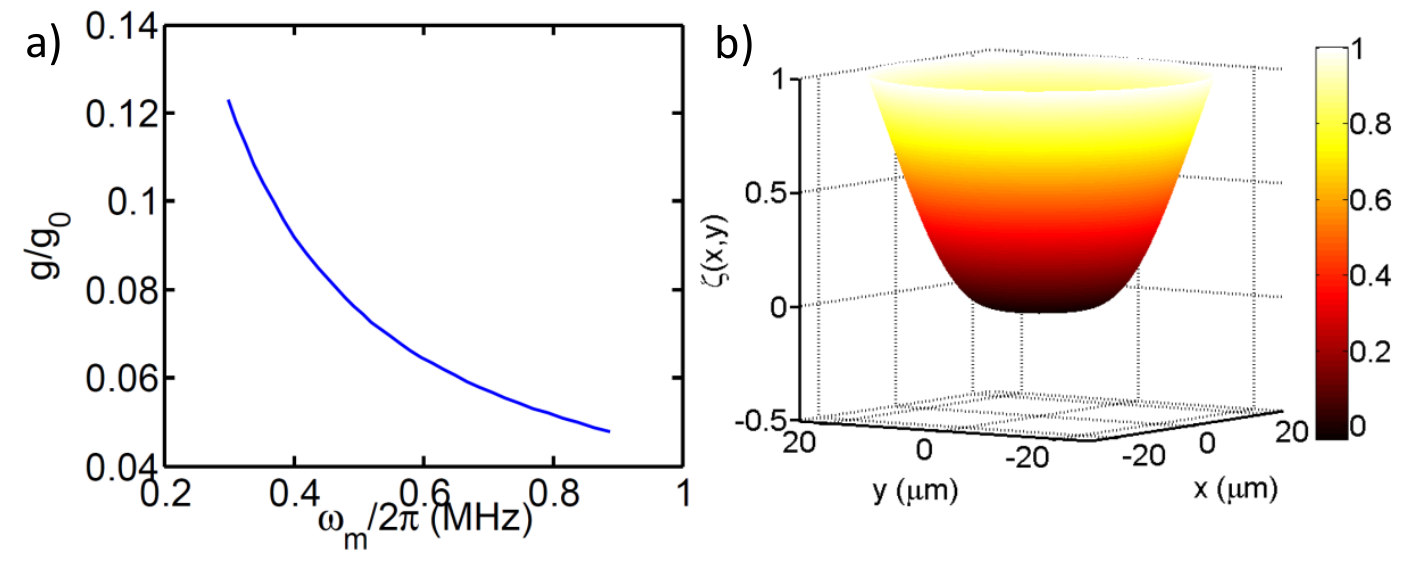}
\end{center}
\caption{a) Opto-mechanical coupling strength $g$ of a trapped
free disk as a function of CM frequency. The coupling strength is
normalized by the value corresponding to rigid (pure CM) motion
$g_0$. A decrease in $g$ for increasing frequency is caused by the
non-uniform optical force pinning the center of the disk in place.
The dimensions for this simulation are $d=30$~nm, $a=25\;\mu$m,
and $w=15\;\mu$m. b) Displacement field $\zeta(x,y)$~(in arbitrary
units) for a free disk of the same dimensions, for a trap
frequency of $\omega_{m}/2\pi=300$~kHz. The displacement field
clearly illustrates the pinning effect created by the optical
forces. \label{fig:g}}
\end{figure}

This effect is illustrated in Fig.~\ref{fig:g} for a free SiN
membrane of thickness $d=30$~nm and $a=25\;\mu$m, interacting with
a beam of waist $w=15\;\mu$m. In this calculation the membrane is
statically trapped, although a similar effect would occur for
sufficiently large dynamical backaction forces as well. In
Fig.~\ref{fig:g}a, we calculate the opto-mechanical coupling
strength $g$~(to another cavity mode with the same beam waist but
which exhibits an intensity gradient at the membrane position) as
a function of the CM frequency, which is varied through the
intensity of the trapping beam. The value of $g$ is calculated
using the expression~\cite{wilson09}
\be
g\propto\,z_{zp}\int\,dx\,dy\,e^{-2(x^2+y^2)/w^2}\zeta(x,y)/\textrm{max}|\zeta|.\label{eq:g}
\ee
In Fig.~\ref{fig:g}a, we have normalized the obtained value of $g$
with the value $g_0$ if the motion were purely CM, where the
displacement field $\zeta$ is constant. At larger frequencies, $g$
dramatically decreases, reflecting the ``pinning'' effect that the
optical force has on the center of the membrane.  This is also
directly seen in Fig.~\ref{fig:g}b, where we plot the displacement
field $\zeta(x,y)$ for a CM frequency $\omega_m/2\pi=300$~kHz.
Incidentally, Eq.~(\ref{eq:g}) also makes it apparent that the
torsional~$(1,0)$ membrane mode described in the main text has
zero opto-mechanical coupling, due to the odd versus even
reflection symmetries of the torsional mode and cavity mode,
respectively.

\section{Tethered membrane inside a Fabry-Perot cavity}

The membrane should be trapped within a Fabry-Perot cavity in
order to enable optical cooling and reach the quantum
regime~\cite{wilson-rae07,marquardt07}. The membrane diffracts and
scatters the cavity light, which introduces two important effects.
First, the optical mode will no longer be Gaussian, and the new
optical mode accommodated by the mirrors and the corresponding
optical forces must be determined. Second, scattering out of the
cavity reduces the cavity finesse, and the associated photon
recoil acts as a stochastic force that heats the membrane motion.
To calculate the cavity modes in the presence of the membrane, we
use a modified Fox-Li propagation technique~\cite{fox68}, as
briefly described below.

Specifically, the electric field is treated within the scalar
paraxial approximation, and thus it is completely described by its
transverse profile $E(x,y)$. This approximation is justified by
noting that the disk should primarily diffract light at small
angles $\theta{\lesssim}(ka)^{-1}$ around the $z$-axis, where
$k=2\pi/\lambda$ is the optical wavevector. Within this
approximation, free propagation over a distance $z$ is accounted
for by a phase shift in the Fourier transform of the field
profile,
$\tilde{E}(k_x,k_y){\rightarrow}e^{ikz-i(k_x^2+k_y^2)z/(2k)}\tilde{E}(k_x,k_y)$.
In our case, we are interested in systems with rotational
symmetry, and thus the transforms are implemented through the
quasi-discrete Hankel transform described in Ref.~\cite{yu98}.
Reflection off of a circular mirror with radius of curvature $R_c$
and reflectance $R_m$ is characterized by the real-space
transformation
$E(x,y){\rightarrow}\sqrt{R_m}E(x,y)\exp\left(2ik(R_c-\sqrt{R_c^2-(x^2+y^2)})\right)$.
Similarly, at the membrane location, the wave front experiences
reflection and transmission amplitudes $r(x,y)$, $t(x,y)$,
respectively, which multiply the incident real-space wave front
there. For the case of a non-uniform disk, we approximate
$r(x,y),t(x,y)$ with the formulas for an infinite thin dielectric
sheet of uniform thickness, replacing the uniform thickness
$d{\rightarrow}d(x,y)$ with the local thickness. Note that an
initial wave front incident on the membrane thus splits into two
wave fronts~(a reflected and transmitted field), and we keep track
of the multiple scattered fields to all orders in order to
calculate the field buildup or cavity eigenmodes. In contrast, the
original technique of Ref.~\cite{fox68} only accounts for
transmission. Thus, our approach properly captures the effects of
the reflected amplitude and back-scattered angle. Furthermore, our
modified technique reveals specifically at what frequencies
resonances should occur.

\begin{figure}[p]
\begin{center}
\includegraphics[width=17cm]{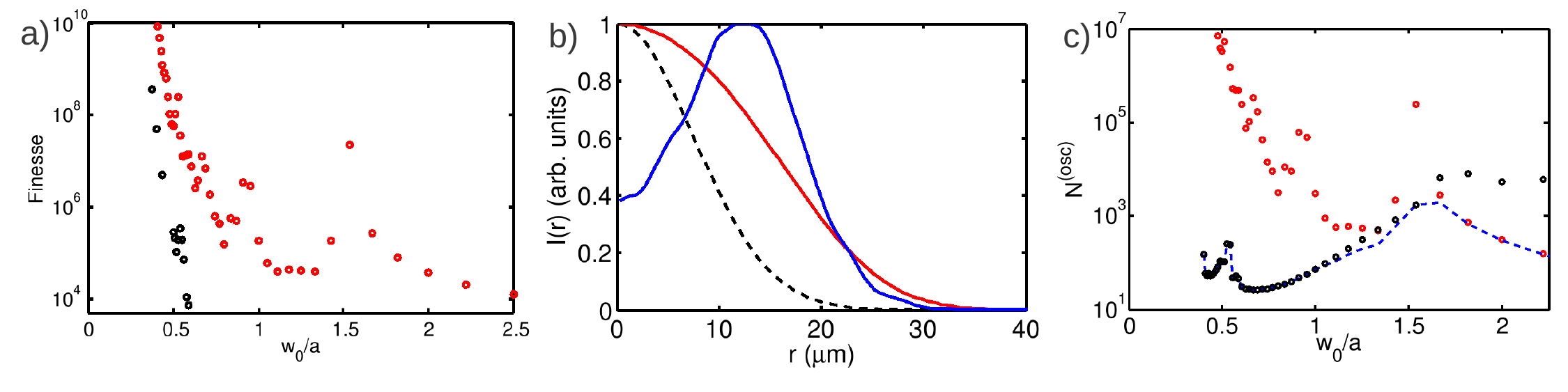}
\end{center}
\caption{a) Membrane-limited finesse
$F_{\footnotesize\textrm{mem}}$ of a Fabry-Perot cavity with a
membrane in the middle. The black circles correspond to a flat
membrane of uniform thickness $d=30$~nm, while the red circles
correspond to an apodized membrane with maximum thickness
$d_0=30$~nm. The finesse is plotted as a function of the ratio of
the empty-cavity beam waist $w_0$ to the membrane radius $a$. The
cavity parameters are chosen such that $w_0=15\;\mu$m. b)
Intensity profiles~(in arbitrary units) of cavity mode in the
presence of an apodized membrane. The intensity profile is
evaluated halfway between the two cavity mirrors. The cavity and
membrane parameters are provided in the main text. The blue and
red curves correspond to disk radii $a=w_0$ and $a=2.5w_0$,
respectively, while the dashed black curve is the Gaussian
intensity profile for an empty cavity. c) The number of coherent
oscillations of the CM motion of an apodized disk due to
thermoelastic
damping~($N^{(\footnotesize\textrm{osc})}_{\footnotesize\textrm{th}}$,
black cirlces) and recoil heating
($N^{(\footnotesize\textrm{osc})}_{\footnotesize\textrm{sc}}$, red
circles), as a function of disk radius and for fixed
$w_0=15\;\mu$m. Also plotted is the total number of coherent
oscillations
($N^{(\footnotesize\textrm{osc})}_{\footnotesize\textrm{tot}}$,
blue dashed curve), which is given by the sum in parallel of the
individual contributions.\label{fig:cavity}}
\end{figure}

We now discuss the effect of the membrane on the cavity finesse.
As realistic parameters, we consider a membrane placed
symmetrically in the center of an optical cavity of length
$L=1.99$~cm with spherical mirrors having radii of curvature
$R_c=1$~cm and perfect reflectivity~(such that the entire cavity
linewidth $\kappa$ is attributable to scattering from the
membrane). The transverse extent of the spherical mirror surfaces
is $r_{m}=0.95$~mm, \textit{i.e.}, all portions of the beam front
with $x^2+y^2>r_m^2$ are scattered out and set to zero upon
reflection at the mirror. An empty cavity in this configuration
yields a Gaussian mode of waist $w_0{\approx}15\;\mu$m in the
center. In Fig.~\ref{fig:cavity}a, we plot the membrane-limited
cavity finesse
$F_{\footnotesize\textrm{mem}}\equiv{\pi}c/{\kappa}L$ for a
membrane of uniform thickness $d=30$~nm and varying radius
$a$~(black circles). Clearly, cavity losses are negligible when
the nominal waist is small compared to the disk radius,
$w_0/a{\lesssim}1$. In the regime $w_0/a{\gtrsim}1$, however, the
finesse rapidly drops, which is attributable to scattering by the
hard edges of the disk. This effect is strongly reduced by
``softening'' or apodizing the disk edge~\cite{born00}. In
Fig.~\ref{fig:cavity}a~(red circles), we also plot the finesse for
a membrane whose thickness $d(r)=d_0(1-(r/a)^2)^2$ tapers down to
zero at the edge, where $d_0=30$~nm is the maximum thickness.
Remarkably, the apodization can improve the cavity finesse by
several orders of magnitude. The modification of the cavity modes
by the membrane is illustrated in Fig.~\ref{fig:cavity}b, where we
plot the transverse profile at the membrane position for some
representative apodized disk sizes.

We find the CM eigenmodes of the apodized disk by using
Eq.~(\ref{eq:waveeqn}), where the optical potential $\omegaopt(r)$
is now evaluated using the modified cavity mode profiles. The
thermoelastic limit is subsequently calculated using Eqs.~(2)
and~(3) in the main text. In Fig.~\ref{fig:cavity}c, the number of
oscillations
$N^{(\footnotesize\textrm{osc})}_{\footnotesize\textrm{th}}$ due
to thermoelastic damping is plotted~(in black). Here the
circulating intra-cavity power is chosen such that the CM
frequency is fixed at $\omega_m/2\pi=0.5$~MHz. We next consider
the effect of photon recoil heating. We assume that each scattered
photon contributes the maximum possible momentum kick of
${\hbar}k$ along the $z$-axis, giving rise to a momentum diffusion
process
$d\avg{p_z^2}/dt=({\hbar}k)^{2}R_{\footnotesize\textrm{sc}}$~\cite{chang10a},
where $R_{\footnotesize\textrm{sc}}$ is the photon scattering
rate. Converting this expression into a jump rate, it can be shown
that the number of coherent oscillations before a jump in the
phonon number can be written as
\be
N^{(\footnotesize\textrm{osc})}_{\footnotesize\textrm{sc}}=\frac{1}{2\pi}\frac{V}{V_c}\frac{\omega_0}{\kappa}\frac{\omega_m^2}{k^{2}I_{\footnotesize\textrm{max}}/{\rho}c}.\label{eq:recoil}
\ee
Here $I_{\footnotesize\textrm{max}}$ is the maximum cavity
intensity~(which in general does not need to be at the center of
the membrane due to mode distortion), $\omega_0=ck$, $V$ is the
volume of the disk, and $V_c$ is the cavity mode volume. Assuming
that the cavity mode is not significantly distorted and that the
beam waist $w_0{\gtrsim}a$ such that the entire membrane
experiences the optical force, one can approximate
$\frac{\omega_m^2}{k^{2}I_{\footnotesize\textrm{max}}/{\rho}c}{\sim}1$
and $V_c{\sim}{\pi}w_0^{2}L/4$. In this case the number of
coherent oscillations scales roughly as
$N^{(\footnotesize\textrm{osc})}_{\footnotesize\textrm{sc}}{\sim}\frac{kV}{w_0^2}F_{\footnotesize\textrm{mem}}$.
Note that this result is purely geometric in nature and also
scales directly with the cavity finesse~(which itself depends on
$V$). In Fig.~\ref{fig:cavity}c, we plot
$N^{(\footnotesize\textrm{osc})}_{\footnotesize\textrm{sc}}$ for
the apodized disk~(in red). Combining the effects of thermoelastic
damping and recoil heating, the total number of coherent
oscillations is given by
$N^{(\footnotesize\textrm{osc})}_{\footnotesize\textrm{tot}}=(N^{(\footnotesize\textrm{osc})-1}_{\footnotesize\textrm{sc}}+N^{(\footnotesize\textrm{osc})-1}_{\footnotesize\textrm{th}})^{-1}$~(blue
curve). It can be seen that an apodized disk of radius
$r{\sim}9\;\mu$m can support a coherence time of
$N^{(\footnotesize\textrm{osc})}_{\footnotesize\textrm{tot}}{\sim}2000$
in a room temperature environment.

\bibliographystyle{apsrev}
\bibliography{../bibsalpha}

\end{document}